# POTENTIAL AND LIMITATIONS OF THE ARCHAEO-GEOPHYSICAL TECHNIQUES

*Yavor Shopov, Diana Stoykova, Antoniya Petrova, Valentin Vasilev, Ludmil Tsankov*

*Archaeological Geophysics Lab., Dept. of Physics, Sofia University, James Baurchier 5, 1164 Sofia; YYShopov@phys.uni-sofia.bg*

**ABSTRACT.** This work demonstrates the potential and the limitations of archaeo-geophysical techniques available at the Archaeological Geophysics Laboratory of the Department of Physics at the University of Sofia with various case studies in natural and artificial environment. Special attention is focused on GPR which is the most powerful archaeogeophysical technique This laboratory is the only one in Bulgaria, which develops new geophysical techniques and equipment for survey of archaeological sites and their dating.

## Introduction

The Archaeological Geophysics Lab of the Department of Physics at Sofia University is the only one in Bulgaria which develops new geophysical methods and equipment for study of archaeological sites and their dating (Shopov et al., 1993; Dermendjiev et al., 1996). This lab has equipment and specialists that use a broad range of archaeogeophysical methods. Here, the possibilities of these techniques to solve various archaeological tasks are demonstrated.

## Archaeogeophysical methods

This lab uses the following archaeogeophysical methods for exploration and non-destructive investigation of archaeological sites:

## I. Radar Methods

**Ground penetrating radar (GPR)**. This method was developed by NASA to study the lunar ground. The introduction of this space technology to archaeology makes GPR the most powerful archaeogeophysical technique (Conyers, 2004), but the interpretation of GPR data is very complicated and requires very complex data computing. It is the most complicated and complex archaeogeophysical technique. GPR allows the registration of such fine archaeological objects that are hard to see by eye and can be missed during archaeological excavations (Conyers, 2004).

*Advantages:* a. GPR is the only archaeogeophysical method which allows preparation of 2D slices (maps) of underground objects at various depths under the surface without their excavation (Conyers et al., 2004) (Fig. 1);

b. It is the only archaeogeophysical method which allows the preparation of 3D reconstructions of the precise shapes and depths of underground objects (Conyers et al., 2004) (Fig. 2);

c. It allows precise determination of depths of the underground objects under the surface;

d. It allows the visualization of underground objects as radar images in real time during the measurements;

e. It allows simultaneous geophysical exploration and archaeological excavation of the registered anomalies.

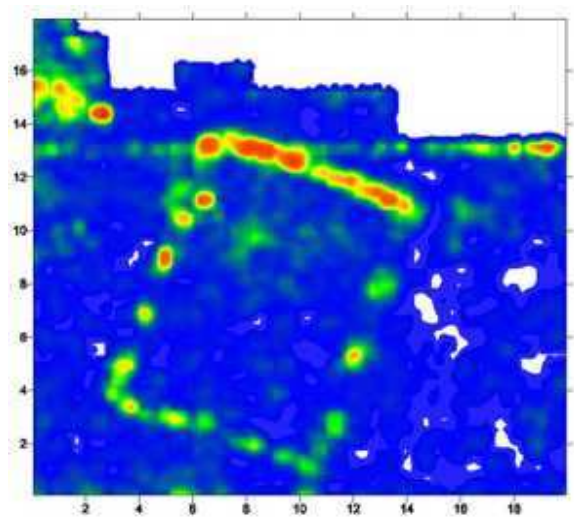

Fig. 1. Amplitude Slice Map 100-150 cm under the surface demonstrating foundations of a building and a possible Roman water line (above) by (Conyers et al., 2004)

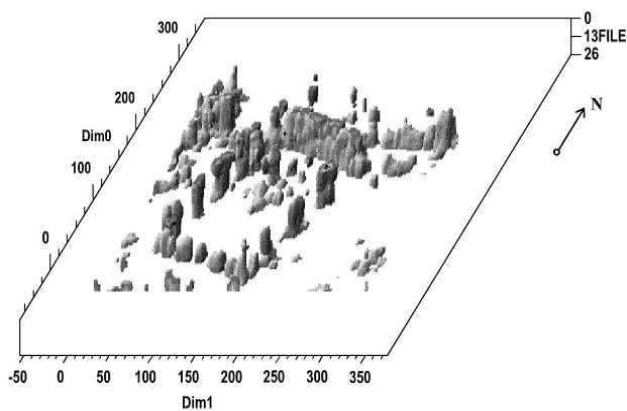

Fig. 2. Three-dimensional rendered surface of the foundations of the building on Fig. 1 constructed from Amplitude Slice Maps like this on Fig.1 from various depths under the surface (Conyers et al., 2004); even separate stones are visible

f. It has the highest resolution from all geophysical techniques;

g. It can be used for scanning of vertical walls and localization of unhomogeneities in it;

h. Registered signal can undergo further computing for extraction of invisible details from the raw scan and graphic display of the results;

i. It allows fast scanning of a large area. It is effective for large scale exploration with high horizontal resolution;

j. It allows connecting of different archaeological excavations by GPR exploration of the space between them;

k. GPR exploration can be done through ice, asphalt, concrete etc. (Archaeological Geophysics Lab website, 2007);

l. On rough terrains step-by-step measurements can be made which allow deeper penetration of the radar signal.

*Disadvantages:* a. The interpretation of the signal is extremely complicated (Conyers, 2004) and requires years of experience of GPR studies of archaeological sites;

b. Very high cost of the equipments;

c. It can not be used in conductive environment (like Sea water) or salty soils;

d. Limited penetration depth which depends on the soil humidity. Usually it varies from 1 meter in wet soil to 17 m in buildings (*Archaeological Geophysics…*, 2007);

e. Archaeological applications of GPR require an expert of very unusual training in specific fields of geophysics, geology and statistical physics. Experience in other GPR applications can not be applied on archaeological sites and such experts can not be easily trained in archaeological applications of GPR.

*GPR applications in archaeology* (*Archaeological Geophysics…*, 2007) is non-destructive localization and mapping of cultural layers in subsequently buried archaeological features: tombs and burials; tunnels, catacombs, wattle-and daub huts and underground channels; building walls; fire places; metal and ceramic artefacts and coatings; cavities and defects in buildings; caves, bunkers, caverns and karst futures; underground reservoirs and buried pipes.

Non-destructive stratification of: sediments, river and lake deposits; soil layers including ancient arable lands; water table; faults and land slides.

Non-destructive study and monitoring of archaeological sites, cultural heritage and underground communications.

## Experimental part
### Calibration Experiments

Large numbers of calibration experiments were made inside the building of Department of Physics of Sofia University (Fig. 3) and adjacent areas with known underground communications (pipes, canals, tunnels, etc.) before the start of the field GPR measurements. They demonstrated that this equipment works perfectly on open ground and inside buildings and visualize all known features of the studied terrains (Fig. 3, 4). It can work up to 17 meters deep in dry environment (Fig. 4). This depth is 70% deeper than the claims of the producer of this GPR unit and antenna.

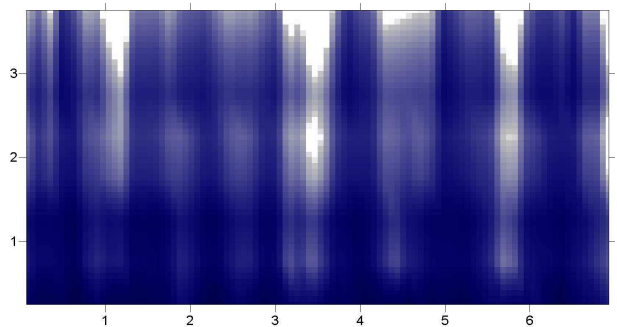
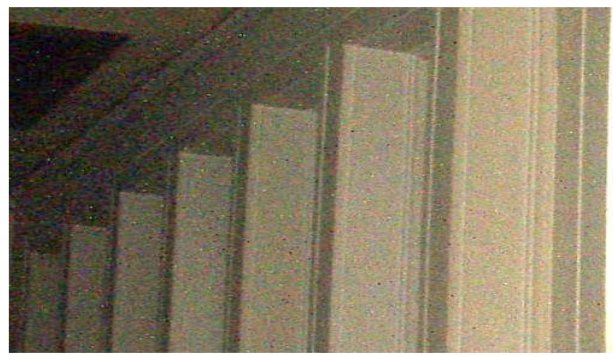

**Fig. 3.** Amplitude Slice Map of the reflection of the radar radiation from the concrete bars on the ceiling of the 4-rd floor measured on 35-62 cm depth through the concrete foundation of the 5[th] floor (Y. Shopov & D. Stoykova); dimensions of X and Y axis are in meters (above); photo of the same concrete bars on the ceiling of the 4-rd floor of building "B" of Dept. of Physics of Sofia University scanned by GPR (down)

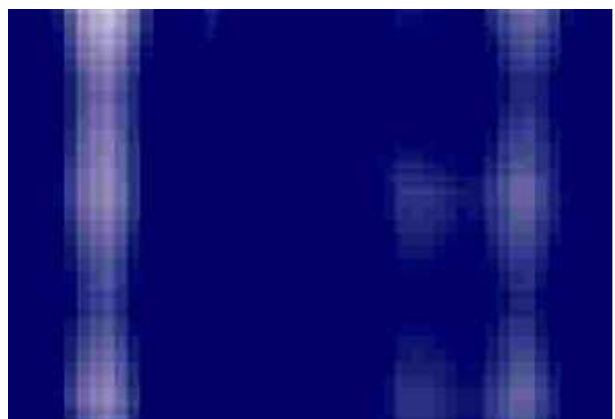
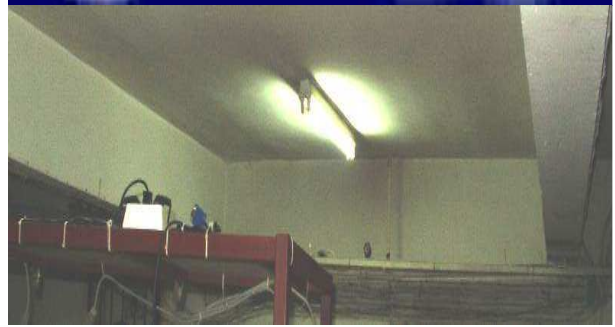

**Fig. 4.** Amplitude Slice Map of the reflection of the radar radiation from the two concrete bars supporting the ceiling of the basement measured on 17.11-17.38 m depth from the 5[th] floor through five concrete foundations with total thickness of 3.25 m (Shopov & Stoykova) (above); photo of the same concrete bars on the ceiling of the basement (-1st floor) of building "B" of Dept. of Physics of Sofia University scanned by GPR (down)

**GPR measurements of Bulgarian archaeological sites**

First GPR measurements on Bulgarian archaeological site (Fig. 5) were made in 2007 in the tomb "Golyamata Kosmatka" (Shopov, in press). 60 scans of the walls and the floor of the tomb were measured with resolution varying from 1.3 to 1.7 cm. Four groups, each of 5 parallel scans, were measured over the walls of the tomb on height from 0 to 250 cm. They were summed in a 3D data base. Then it was sliced to 15 slices (Fig. 6) of 20 nanoseconds (corresponding to a thickness of 75 cm if the radar beam passes through soil and up to 3 m – of it passes through air).

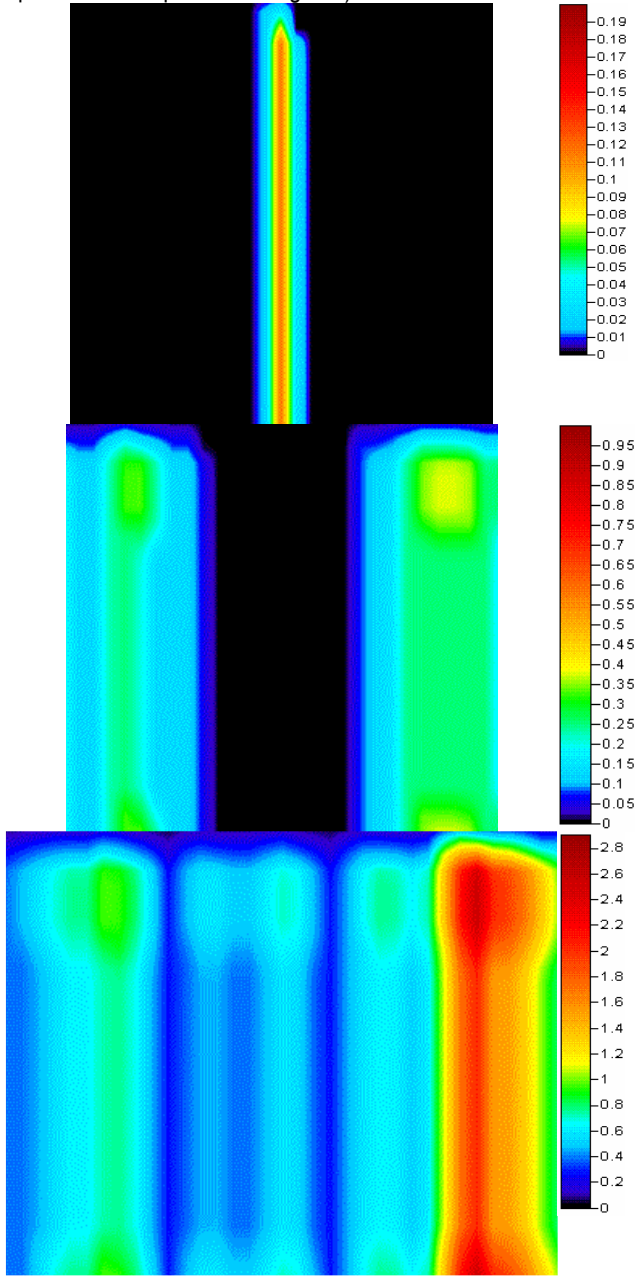

Fig. 5. Vertical Amplitude Slice maps of the intensity of the radar radiation reflected by objects at the "Golyamata Kosmatka" tomb (Fig. 6) measured through the wall of its round camera; a vertical slice 100-150 cm behind the wall of the camera (above) – external wall of another unknown round building is intersected in the middle of the scans; a vertical slice 150-225 cm behind the wall of the camera (middle) – in the beginning and the end of the scans the external walls of the other round building are intersected; (Down) a vertical slice 450-525 cm behind the wall of the camera (below) – in the beginning and the end of the scans external walls of the other round building are intersected, three vertical structures between them can be internal columns (color codes of the intensity of the reflected radar radiation are given to the right)

The obtained slices have resolution of 0.1 m in horizontal, but 0.5 m in vertical direction. The scanned tomb camera is round, so the obtained slices are segments of a circle (Fig. 6). The 2D maps look as prints of cylindrical seals (Fig. 5). They demonstrate that a second unexcavated camera is located behind the west wall of the tomb. It is twice as big as the camera in the excavated tomb.

Scans of the walls of the tomb in the lowest scanning position suggest that the radar radiation penetrates trough homogeneous material (Fig. 7a) at least to 16 m in all directions. The material of the walls is granite. It means that the whole tomb is embedded at least 50 cm deep in a granite square and at least 35 m in diameter. This does not mean that the square is circled. It can be extended in all directions but the radar radiation can not reach its edges. The soil fill of the mound is detected through the granite wall of the tomb (Fig. 7b) everywhere over 50 cm above the floor.

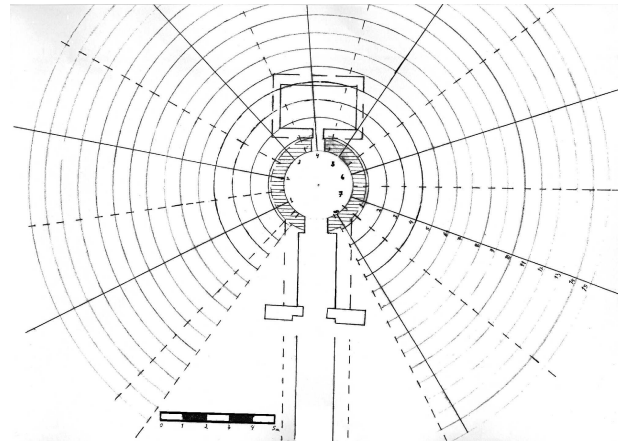

Fig. 6. Scheme of the tomb, distribution of the radar radiation during its scanning and positions of the 15 slices of 20 nanoseconds each (corresponding to a thickness of 75 cm if the radar beam passes through soil but up to 3 m through air). It was unusually complicated because all important scans are vertical due to the great depth of the tomb, which makes it impossible to measure from the surface of the mound by GPR

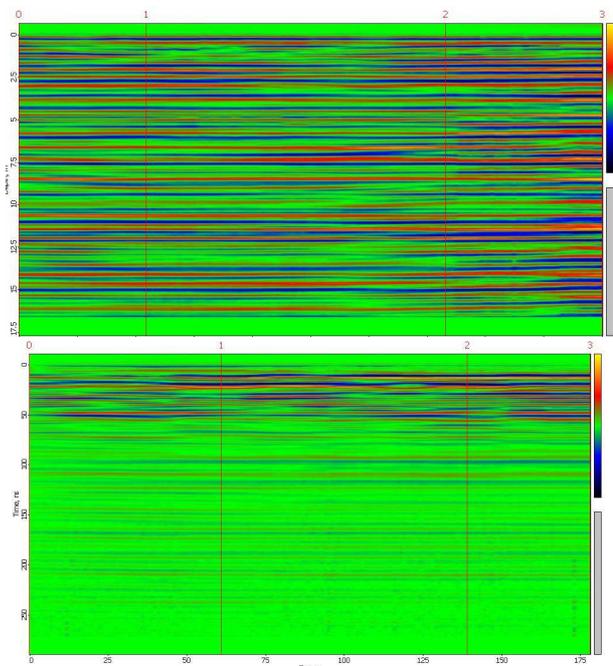

Fig. 7. A scan of the walls of the vestibule of the tomb 0-50 cm above the floor suggesting that the radar radiation penetrates through homogeneous material at least 16 meters in all directions (above); a scan of the walls of the vestibule of the tomb 200- 250 cm above the floor (below) – it demonstrates that the radar radiation penetrates through the granite wall of the tomb in the homogeneous soil fill of the mound outside the tomb wall

**GPR measurements of prehistoric archaeological sites**

Prehistoric sites are the most difficult archaeological objects for archaeogeophysical survey due to lack of metal objects. Most of the artifacts have the same chemical composition and physical properties as the surrounding ground. Especially the stone artifacts have the same properties as the stones around. So GPR is the most appropriate archaeogeophysical technique for survey of Neolithic settlements (Fig. 8) and is the only one usable for survey of Paleolithic sites.

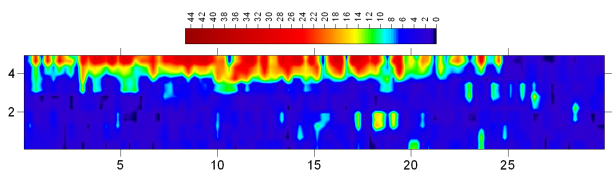

Fig. 8. Amplitude Slice Map, 252-261.5 cm under the surface of an archaeological site in South Bulgaria demonstrating foundations of a possible stone wall (above) of a potential Neolithic building measured by Y. Shopov, A. Petrova, D. Stoykova and V. Vasilev; dimensions of X and Y axis are in meters

GPR is the most suitable geophysical technique to solve many of the tasks of archaeological exploration. Before its development it was considered impossible to locate underground objects like plastics, terracotta, concrete and asphalt. GPR became the main technique for localizing and mapping of non-conductive, non-metal and non-magnetic objects. It can even be used for exploration of under-water objects in fresh water basins (*Archaeological Geophysics…*, 2007).Therefore, in the last years it is the main focus of work of Archaeological Geophysics Lab at Sofia University.

## II. Electrical resistivity methods

**Electrical profiling.** It measures profiles of the electric resistivity (Fig. 9). It allowed the deepest geophysical exploration of a Bulgarian archaeological site at 19 m below the surface (Shopov, 2007) but such measurements can be done even to 40 m depth. The methods is most appropriate for searching of tombs, caves, tunnels or bunkers.

**Vertical electrical probing.** The method detects the same objects as electrical profiling – serving for determination of the depth of the detected anomalies.

**Electrical tomography** (continuous electrical probing). Allows the visualization of anomalies of electric resistivity and of the objects that create it.

Although their great depth of operation these methods are extremely slow, labour-intensive and expensive, and have many limitations and interferences. So now the Archaeological Geophysics Lab has abandoned these methods except the Vertical Electrical Probing which sometimes can help GPR to determine the depth of the detected anomalies.

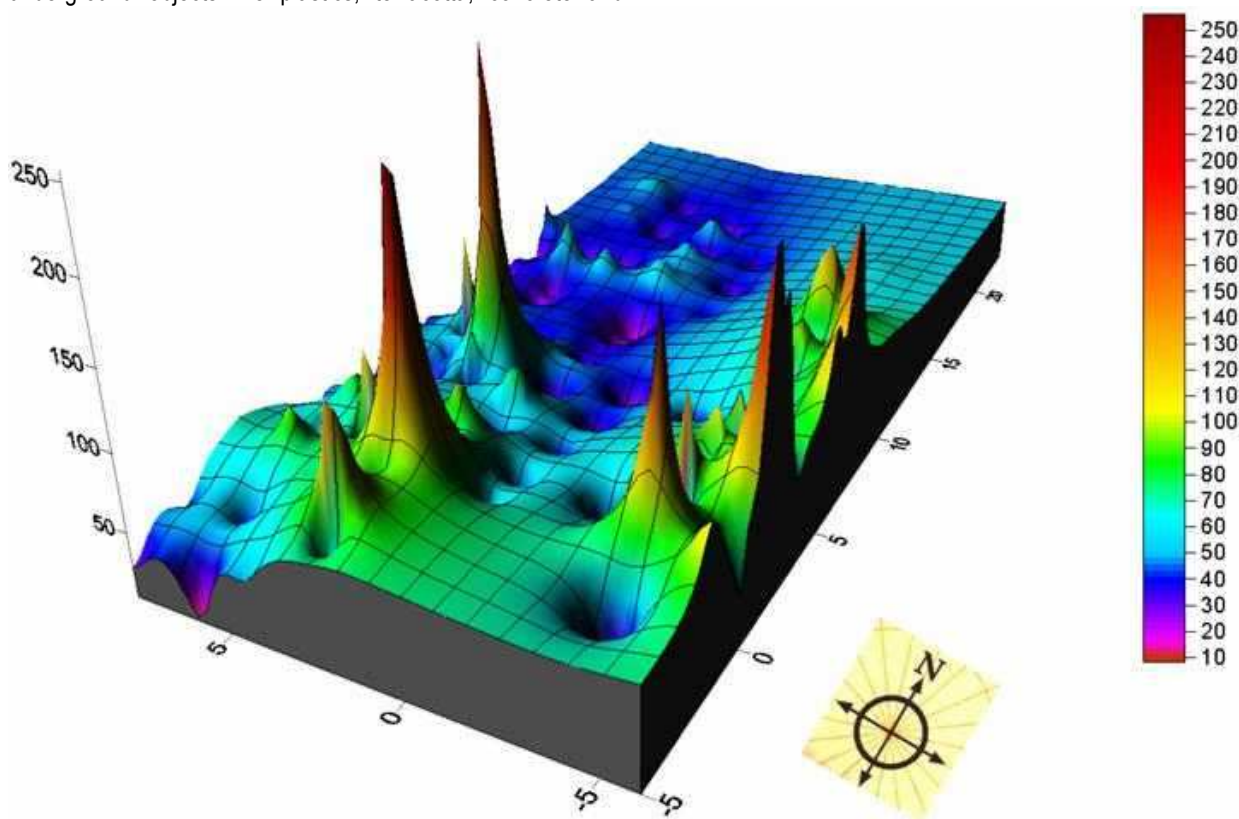

Fig. 9. Map of the electric resistivity of the Omurtag tomb; vertical axis is in units of Omh/m (Shopov, 2007)

**III. Induction methods.** Application as military technologies for location of mines.

**Pulse induction.** Allows localization of large metal objects at depth up to 6 meters. Its equipment emits powerful electromagnetic pulses and measures the inducted current of the underground objects between the pulses (Aittoniemi et al., 1986). It works through walls and stones. It allows very fast scanning and high precession of localization of the objects, but it does not allow precise determination of the depth of the anomalies. Underground cables, rebar or metal networks mask the objects and make its use impossible.

**Electromagnetic induction.** Allows precise localization of small metal objects and determination of the metal by its conductivity (Gardiner, 1967). Works on shallow depth which varies from 0.3 up to 1 m depending on the size of the found object. Its equipment emits electromagnetic field and measures the inducted current of the underground objects passing between its coils. It does not allow the determination of the depth of the anomalies. Underground cables, rebar or metal networks mask objects and make its use impossible.

Due to the limitations of each method in some cases it is necessary to use several methods and different equpment to solve a specific task. All geophysical explorations are non-destructive and harmless for the archaeological features unlike coring which damages the features in some degree.

*Acknowledgements.* We express special thanks to V. Mutafov, A. Koichev and M. Purvin for their help in the terrain measurements, to Assoc. Prof. Dr. D. Gergova, Assoc. Prof. Dr. G. Kitov and Assoc. Prof. Dr. K. Leshtakov for the helpful data, discussions and provision of technical support at their excavation sites. We thank to Hydroloc Ltd. for providing their equipment for field research. We thank Prof. G. Tenchov for the discussions.